# Concurrent transitions in wear rate and surface microstructure in nanocrystalline Ni-W

Jason F. Panzarino [1], Timothy J. Rupert [1,2,*]

[1] Department of Mechanical and Aerospace Engineering, University of California, Irvine, CA 92697, USA

[2] Department of Chemical Engineering and Materials Science, University of California, Irvine, CA 92697, USA

*E-mail: trupert@uci.edu

**Abstract**

Nanocrystalline metals are promising materials for wear-resistant applications due to their superior strength and hardness, but prior work has shown that cyclic loading can lead to coarsening. In this study, scratch wear tests were carried out on nanocrystalline Ni-19 at.% W films with an as-deposited grain size of 3 nm, with systematic characterization performed after different wear cycles. A new gradient nanograined microstructure is observed and a direct connection between wear rate and subsurface microstructure is discovered. A second Ni-W specimen with the same composition and a 45 nm average grain size is produced by annealing the original specimen. Subsequent wear testing shows that an identical subsurface microstructure is produced in this sample, emphasizing the importance of the cross-over in deformation mechanisms for determining the steady-state grain size during wear.

# 1. Introduction

Nanocrystalline metals are known to exhibit high strength [1, 2] and fatigue resistance [3, 4], which together suggest that these materials will have excellent wear resistance. These enhanced mechanical properties, combined with the ability to tune the grain size of these materials [5-7], have driven their implementation as protective coatings with friction coefficients and wear rates that are dramatically lower than their coarse-grained counterparts [8-11]. However, wear is also a relatively extreme deformation scenario, where the surface is potentially exposed to high stresses, strain rates, mechanical mixing, and frictional heating, meaning the microstructure near the surface can be dynamic during testing. Coarse-grained metals experience the formation of cellular dislocation sub-structures [12, 13], mechanically-mixed layers [14, 15], or nanocrystalline tribolayers [16-22]. However, the creation of such structures near the contact surface requires dislocation storage and tangling mechanisms, which should not be active in nanocrystalline metals.

Nanocrystalline metals deform through unique grain boundary-mediated mechanisms such as grain boundary migration and grain rotation [23-25], as well as dislocation nucleation and pinning at grain boundary sites [26, 27]. The operation of new plasticity mechanisms means that any sub-surface evolution of the microstructure during wear should be novel as well. Recent experimental studies in this area have begun to uncover the range of possible structures that can be found after wear. For example, Hanlon et al. [28] found evidence of coarsening near the contact surface in nanocrystalline Ni with an average grain size ($d$) of 30 nm, using top-down imaging with a focused ion beam (FIB) microscope to visualize the grain structure. Mechanically-driven grain growth of nanocrystalline materials has also been reported for tensile [29] and fatigue experiments [30]. This behavior appears to be driven by high shear stresses [31], but can be dramatically affected by the impurity content of the material [32]. Qi et al. [33] observed that sliding wear caused coarsening as well as grain elongation and texturing in Ni with $d = 10$ nm. Layered substructures have also been discovered beneath wear surfaces in materials that were initially nanocrystalline. Prasad et al. [34] were able to confirm the formation of a nanocrystalline surface layer atop a coarsened substructure in worn Ni films which initially contained 20-100 nm grains. Padilla et al. [35] extrapolated upon this finding and discovered that the permanence of the top nanocrystalline tribolayer is dependent on the friction coefficient, which will dictate micro-cracking and delamination of the layer.

Unfortunately, the majority of these studies do not focus on the smallest nanocrystalline grain sizes, where grain boundary mediated deformation physics should dominate plasticity and the most extreme structural evolution is expected. However, a few select studies do exist. Rupert and Schuh [36] analyzed Ni-W films with average grain sizes from 3 to 47 nm to understand sliding wear across the breakdown in Hall-Petch scaling. These authors discovered deviations from the Archard scaling law which could be attributed to grain boundary relaxation strengthening, a phenomenon characteristic of as-deposited nanocrystalline materials with average grain sizes of only a few nanometers. Ball-on-disk wear of $d = 3$ nm Ni-W resulted in post-deformed surface structures containing a well-defined layer of grain growth, while the tests involving larger starting grain sizes revealed little to no surface structure evolution. Argibay et al. [37] analyzed similar Ni-W coatings with $d \sim 5$ nm and suggested that the thickness and average grain size within the evolved layer is a function of applied stress, correlating to the friction behavior. More recently, Argibay et al. [38] developed a model which predicts either coarsening or refinement in surface tribolayers as a function of the equilibrium dislocation splitting distance ($r_e$), with coarsening occurring when $d < 2r_e$ and refinement when $d > 2r_e$. The exact grain size to which an evolved surface layer will eventually converge is then dependent on the applied surface stress. In a subsequent study on pure Cu sliding contacts at cryogenic temperatures, Chandross et al. [39] presented indirect experimental validation by calculating the expected equilibrium grain size as a function of friction coefficient, relating surface friction to constitutive models of grain size dependent strength. These studies clearly highlight the fact that the finest nanocrystalline microstructures are dynamic during wear, allowing for the formation of interesting tribolayers. However, an understanding of how these layers develop is lacking since microstructural characterization was only performed pre- and post-deformation. A detailed connection of various wear substructures and the instantaneous wear rates is also missing.

In this study, we analyze subsurface evolution of Ni-W with an extremely small starting grain size ($d$ = 3 nm) during abrasive wear using a nanoindenter-based scratch method. Wear volume is tracked during testing, to provide a guide for characterization of the surface microstructure. We find three distinct regions of wear behavior, characterized by wear rates that each vary by an order of magnitude. Due to the micrometer-sized geometry of the scratch test, the full wear track is able to be incorporated into a single transmission electron microscopy (TEM) lamella. First, the grain structure remains static while the surface of the material deforms to match the indenter shape. Next, a gradient nanograined structure develops, with a smooth transition from coarser surface grains to a fine nanocrystalline grain structure identical to the as-deposited material. Finally, a discrete layer of coarsened material with an average grain size of ~10 nm is found. Finally, we anneal an identical Ni-W sample, to create a film with the same chemistry but a significantly larger grain size, and repeat our wear experiments. We find that nanocrystalline materials coarsen or refine to a single grain size that can be connected to the transition from purely boundary-based to dislocation-dominated deformation physics.

## 2. Methods

A nanocrystalline Ni-W alloy was created using the pulsed electrodeposition technique developed by Detor and Schuh [7]. The film was deposited onto a Ni-200 substrate that had been mechanically polished to below 1 µm surface roughness using diamond suspension, with the nanocrystalline films being ~60 µm thick. A Pt mesh counter-electrode was used in a continuously stirred, constant temperature plating bath free of organic refiners described by Yamasaki et al. [40]. After deposition, the surface was again polished to below 1 µm roughness in order to achieve a smooth and level surface for subsequent wear experiments. X-ray diffraction (XRD) analysis of the final film was performed with a Rigaku SmartLab X-ray diffractometer using a Cu Kα radiation source operated at 40 kV and 44 mA. An average grain size of 3 nm was measured using the Scherrer equation after correcting for instrumental broadening. This average grain size was verified using a FEI/Philips CM-20 TEM microscope operating at 200 kV. A global film composition of Ni-19 at.% W was measured in an FEI Quanta dual-beam scanning electron microscope (SEM)/FIB using energy-dispersive X-ray spectroscopy (EDS) operating at 30 kV. To create a larger grain size specimen with the same composition, another piece of the sample described above was annealed at 700 °C for 24 h. Annealing was carried out in a tube furnace under flowing Ar and all heat treated samples were also re-polished to remove any oxide layer that could have potentially formed. XRD and TEM found that the average grain size had increased to 45 nm after this annealing treatment. This annealing procedure will also relax any nonequilibrium grain boundaries present in the as-deposited state [41] which may lead to some increase in the hardness and wear resistance of this sample. Unfortunately, full relaxation of the 3 nm sample would require simultaneous grain coarsening [36], meaning a fully relaxed sample with the finest grain size is impossible to make. In the end, we choose to keep the composition constant between the two starting grain sizes, despite the existence of some variation in the grain boundary relaxation state. We believe that this is acceptable because the main goal of this part of the study is to understand how different deformation mechanisms (grain rotation/sliding vs. dislocation-based deformation) evolve the subsurface microstructure.

Scratch wear testing of the films was performed using an Agilent G200 Nano Indenter fitted with a 60° diamond conical tip with a 5 µm radius. Figure 1(a) shows an example of a wear scar, where the probe started at the left end of the scar. Each wear cycle involves the application of a constant normal load of 10 mN while traversing the surface of the material in both the pass and return directions (i.e., each cycle hits a given surface point twice). This process is repeated 10 times before a transverse profiling pass is performed. For the profiling pass, which can be used to monitor wear volume during testing, the indenter tip performs a cross-profile of the scratch trench by moving perpendicular to the sliding direction using a much lighter load. A profiling track is visible in Figure 1(a) as the vertical line that passes through the wear track. A high profiling load was used in this case for instructive purposes, but a much lower profiling load of 0.05 µN (low enough to not visibly mark the surface), was used for all tests that are reported here. This profiling load is near the lower limit of the G200 system in order to avoid any additional deformation of the track during profiling. After profiling the trench geometry, the probe then returns to the starting location and the entire procedure is repeated until a specified number of wear cycles have been executed. The wear scar shown in Figure 1(a) experienced 1000 wear cycles, meaning that 100 cross-profiles were taken during the experiment. The large piece of debris which can be seen on the wear path is a piece of dust, which conveniently provides a size reference for the geometry of our wear experiments.

Because surface profiling is performed at regular intervals during the experiment, material removal and pileup morphology can be tracked as a function of cycle number. Figure 1(b) shows a two-dimensional cross section of a wear track in the as-prepared state and after 10 sliding cycles. Total wear volume is determined by integrating the area above and below the initial surface topography and then multiply this area by the length of the scratch (200 µm). This measurement thus looks at total displacement of material during the scratching process. Integration of the worn area in the two-dimensional cross profile is illustrated in Figure 1(b) by the blue hatched area above and below the un-deformed film surface. Emerging trends in the instantaneous wear rate were used to influence future testing runs and pinpoint cycles of interest for subsequent subsurface microstructural characterization. Once an interesting trend in wear rate has been identified, future wear tests could then be truncated after certain cycling levels so that TEM lamella from the wear track could be extracted using the FIB lift-out technique [42]. The sample surface was protected from the ion beam during lift-out and thinning with a Pt film that was first deposited under the electron beam and then under an ion beam, as shown in Figure 2(a). TEM samples were removed from the surface (Figure 2(b)) and mounted at the top of V-shaped posts on TEM grids in order to help prevent bending or snapping which can occur when the samples are thinned to 50 nm or below. During the sample thinning, the current of the Ga ion beam was iteratively decreased to a final setting ≤5 kV, which ensures removal of the damage layer left by the earlier steps in the sample thinning process. A mounted TEM specimen which is midway through the thinning procedure can be seen in Figure 2(c). The cross-section of the wear scar is clearly visible in this image below the lighter protective Pt layer. The full geometry of the wear scar is captured within the TEM lamella, which allows for characterization of microstructure in different regions within the wear scar.

## 3. Results and Discussion

*3.1. Scratch Wear Experiments on d = 3 nm Sample*

To begin, scratch wear tests to 1000 cycles were performed, with wear volume as a function of cycle displayed in Figure 3(a) for five such experiments. Wear rate, or the rate of change of this data, appears to be very high at the beginning of each experiment but become smaller as the tests continue. To more clearly visualize trends in the data and allow for detailed analysis, the wear volume at a given cycle number was averaged across the five tests. Figure 3(b) shows the averaged wear volume data, where distinct wear rate trends are easy to identify. Three possible wear regimes are labeled with dashed lines, with the linear equations for each of the trend lines displayed in the figure. It is important to note that we do not mean to suggest any physical meaning to this linear fit or the associated equation, but instead simply use these three trends lines to identify different regions of interest. During the beginning of the test, the wear removal rate is highest as indicated by the steep slope of the red trend line. The rate then sharply decreases by roughly an order of magnitude after ~20 cycles and then roughly another order of magnitude after ~200 cycles as indicated by the blue and green lines, respectively. In order to uncover any connections between these wear regimes and the underlying microstructural evolution that takes place below the scratch tip, additional scratch experiments were stopped at cycle numbers slightly past the shifts in the wear evolution trends of Figure 3(b) in order to characterize the microstructure at these points of interest. FIB-prepared TEM lamella were made perpendicular to the scratch direction for experiments after 1, 30, 250, and 1000 cycles, as indicated by the gold stars in Figure 3(b).

The surface morphologies of the wear scars were first analyzed using SEM for evidence of the wear mechanisms taking place during the scratch experiments. Figure 4 displays surface images from 1, 30, 250, and 1000 cycle wear tests. In Figure 4(a), the inset image provides a magnified view of the upper half of the trench and points out several locations where flow localization is caused by inhomogeneous plasticity within the pileup during the first scratch cycle. Flow localization or shear banding has been observed in other experimental studies of the mechanical response of nanocrystalline Ni-W system [43, 44]. Extensive local plastic deformation is required in the first sliding cycle in order for the surface to match the geometry of the probe tip. Another important observation is that there is no evidence of debris, meaning all of the wear volume is material that is displaced during the initial plowing by the conical tip. Schuh et al. [45], suggested that such a blunt tip geometry should impose high hydrostatic pressures on the surface and also reported no evidence of debris for their single cycle scratch experiments that ramped the tip load from 0 – 10 mN. Similar to the smooth trench resulting after 1 cycle, the 30 cycle trench of Figure 4(b) also shows a smooth surface, albeit with a wider wear scar and no visible flow localization at the edges of the track, possibly due to more displaced material which hides such evidence. There is also no identifiable wear debris present along this wear scar.

Figure 4(c) and (d) show morphologies that are visibly different from earlier cycles. While the center of the track is still relatively smooth, visible surface roughness has developed toward the outside of the wear scar. There is also evidence of an additional pileup layer in Figure 4(c) and (d), as indicated by the black arrows. There is no apparent evidence of a transfer layer, likely due to low energy of adhesion between diamond and metal contacts as well as the back-and-forth sliding method used for these experiments. However, wear debris is clearly visible outside of the wear track for the wear experiments at higher cycle numbers. The wear debris particles are not elongated or plate-like, providing visual evidence that this is not a case of delamination wear. In addition, the calculation of specific wear rates [46] for the three consecutive wear stages gives values of $9.7 \times 10^{-4}$ mm$^3$/N-m, $6 \times 10^{-5}$ mm$^3$/N-m, and $1 \times 10^{-5}$ mm$^3$/N-m, all of which are not generally low enough to be typical of delamination wear. Such wear debris was only observed for the 250 and 1000 cycle tests but would not appear in the surface traces shown in Figure 1, indicating that the wear volume measurements may be too low at these higher cycle numbers. Including this material in our integrated wear volume calculations would thus alter the trend line of our data for the green regime in Figure 3(b) and consequently our saturation wear rate of $1 \times 10^{-5}$ mm/N-m. Since the purpose of this study concerns microstructural evolution during wear and not the exact slope of the wear loss curve, we are mainly interested in the material which has undergone repeated plastic deformation without removal from the surface and focus on the materials comprising the main wear track.

### 3.2. Subsurface evolution in d = 3 nm sample

Cross-sectional TEM lamella were cut perpendicular to the scratch direction from tests run to 1, 30, 250, and 1000 cycles, with bright-field TEM images shown in Figure 5. Our TEM analysis captures grain structure evolution surrounding the scratch tip as wear damage accumulates during cycling. Figure 5(a) shows the entire scratch width of a 1 cycle experiment with both pileup regions visible at the corners of the image. The selected area electron diffraction (SAED) pattern shown in the inset of Figure 5(a) has only solid rings that can be indexed as those from a face centered cubic polycrystalline material. The grain structure below the wear surface is not noticeably altered from that of the as-deposited microstructure, although a few slightly enlarged grains right at the surface were found and are indicated by the white arrows in Figure 5(b). As the number of sliding cycles is increased to 30, significantly more evolution is observed as shown in Figure 5(c) and (d). While grain growth has occurred, there is a gradient in the average grain size which changes as a function of both depth and distance from the center of the scratch tip. The magnified image in Figure 5(d) more clearly shows this gradient as one moves away from the surface.

Rupert and Schuh [36] calculated that the stresses during their pin-on-disk wear experiments should be the highest just beneath the traversing spherical probe through the use of the contact mechanics equations from Hamilton [47]. Although these equations assumed a perfectly spherical probe, our probe geometry is similar since our conical indenter tip is used at fairly shallow indentation depths. In addition, nanocrystalline metals have been reported to coarsen through stress-driven grain boundary migration [31, 48, 49], with such migration usually resulting in discontinuous grain growth [29]. Recent molecular dynamics simulations by Panzarino et al. [50] also verify that grain rotation and coalescence can contribute to discontinuous grain growth during cyclic loading and that the magnitude of such coarsening increases with increasing number of applied loading cycles. Experimental observation of grain growth and its dependence on shear stress was also reported by Rupert et al. [31], with higher stresses leading to more evolution. It is logical then that this gradient structure has developed since the subsurface von Mises equivalent stress should decrease with depth.

Analysis of the 250 cycle sample in Figures 5(e) and (f) shows that the gradient has disappeared by this point and the damaged material has formed a well-defined coarsened layer. This layer extends ~270 nm into the film when measured at the center of the conical scratch tip and the average grain size in this layer was calculated to be 11 nm. The grain size distribution is relatively broad, with grains from 3 to 40 nm found in this layer. The white arrow in Figure 5(e) shows where the pileup of worn material has begun to separate from the bulk surface, which aligns with the prior observation of wear debris for the tests consisting of 250 or more cycles. As the number of cycles is increased to 1000 (Figure 5(g) and (h)), the subsurface microstructure appears to be relatively constant. The coarsened layer extends ~300 nm below the contact surface and the overall shape of the layer is geometrically similar. The SAED patterns taken from both the 250 and 1000 cycle samples show more discrete spots, compared to the continuous rings found in the as-deposited materials and after 1 cycle. However, no clear preferred texture could be indexed. In general, Figure 5(h) shows a grain structure that is qualitatively similar to that shown in Figure 5(f) and the calculated grain size of the evolved layer in the 1000 cycle sample is 10 nm, within measurement error of the grain size observed in the 250 cycle sample. Evidence that the grain growth is discontinuous can be seen as small grains characteristic of the as-deposited sample which remain in the microstructures of both the 250 and 1000 cycle sample. This provides further evidence that grain boundary mediated mechanisms are taking place to allow for the migration of certain favorable boundaries.

Most striking in the analysis shown here is the fact that these three stages of subsurface structural evolution ((1) material displacement and conformity to the scratch tip, (2) formation of a gradient nanograined structured, and (3) saturation to a distinct coarsened layer) directly correlate with the three wear regimes highlighted in Figure 3(b). The fact that the growth layer evolved from a gradient structure can be explained by considering the stress gradient underneath the sliding contact. The magnitude of plastic strain which occurs will scale with the stress level experienced by each region of material. Stresses are highest at the surface and drop off as one moves into the film. Grain growth therefore occurs most rapidly where the stresses are highest at the surface, but gradually stops as the grains grow. Grain boundary migration, grain rotation, and grain sliding are only active at extremely small grain sizes, shutting off above $d \sim$ 10-20 nm and being replaced by dislocation-based mechanisms [51-53]. This average grain size can be called the "cross-over grain size," since it represents a microstructural length scale where the deformation mechanisms responsible for plastic strain fundamentally change. Regions below the surface, which experience stresses that are lower but still high enough to cause yielding, will coarsen much more slowly and require more sliding cycles to reach the cross-over grain size. The similar thickness and grain size of the evolved layer at higher cycle numbers suggests that a steady-state microstructure has been reached underneath the sliding contact. Material is worn away and the previously unaffected material (now closer to the surface than before) will undergo coarsening, but the thickness of the layer does not dramatically change.

Evolution of nanocrystalline grain structure and the accompanying grain boundary network during cyclic loading has been shown by Panzarino and coworkers [50, 54] for molecular dynamics simulations of nanocrystalline Al with grain sizes similar to the Ni-W studied here. In addition, recent in situ synchrotron X-ray diffraction experiments by Furnish et al. [30] demonstrated abnormal grain growth during tensile fatigue in Ni-Fe with $d$ = 18 nm, with evidence that grain boundary mediated plasticity was the driver for such growth. Regarding wear in particular, Argibay et al. [37] also reported on a coarsened tribolayer in nanocrystalline Ni-W that was visibly similar to our own, with coarsening signifying a shift to higher friction coefficients during sliding. These authors proposed a model suggesting that low-friction sliding will evolve microstructure slowly toward a steady-state grain size through grain boundary mediated mechanisms. Higher stresses would achieve this coarsened state much more quickly, thus signaling the crossover to higher friction coefficients. The layers observed by Argibay et al. were more clearly defined at higher loads, caused by the increased Hertzian contact stress below their larger millimeter-sized sapphire sliding contact. The geometry of our micron-scale probe allows for very high contact stresses (Hertzian contact mechanics calculations give values in the GPa-range that are well above the expected yield strength), which provides an explanation for why we observe a well-defined saturation grain size. More recently, Argibay et al. [38] extrapolated on their own previous work and developed a model which predicts either coarsening or refinement below the sliding contact, stating that the crossover grain size is simply twice the equilibrium dislocation splitting distance ($r_e$) of the material. Most notable is the fact that our findings here appear to validate their model quite well. According to their model, a low friction contact combined with a starting grain size below the $2r_e$ threshold should evolve the subsurface microstructure to a larger equilibrium grain size. For pure Ni, $2r_e$ is predicted to be 9.8 nm which would mean that we should indeed expect to observe coarsening in our 3 nm samples.

These three stages of evolution have a direct consequence for the wear response of our Ni-W film. The wear rates in Figure 3(b) decrease as the stages of evolution occur, which we can trace to Archard-type scaling that predicts that wear volume is inversely related to hardness. Although grain growth is occurring below the tip, the amount of growth is small and the change in grain size would not be expected to alter hardness since this property is independent of grain size for such small values [7]. However, grain growth can be accompanied by relaxation of nonequilibrium grain boundary structure, which significantly increases hardness [43]. Panzarino et al. [50] also showed that strengthening can occur during cyclic deformation as a result of the formation of lower energy boundary configurations. For the gradient nanograined structure, only the coarsened grains near the surface have boundaries with relaxed structure, while the grains below remain close to the as-deposited state. However, once the saturated coarsened layer is formed, the film has a fully hardened layer at the surface.

Additional evidence that these processes are ongoing throughout wear deformation can be found by investigating the near surface microstructure near the edges of the scratch trench. For example, while the evolved layer is well-defined near the center of the trench for the 250 and 1000 cycle samples, remnants of a gradient near the edges of the wear scar can be seen in Figure 6(a) and (b). This occurs because the material near the edges has been pushed away from the center and has therefore been subjected to a smaller number of cycles at the highest loads. Flow localization in the form of shear banding can also be seen in Figure 6(b), which allows for the displacement of the pileup material. Inhomogeneous plasticity, evidenced by the curved line of localized grain growth, is pointed out by the red arrow in Figure 6(b). Khalajhedayati and Rupert [44] observed similar shear bands in the pileup of nanocrystalline Ni-W during nanoindentation.

*3.4. Subsurface evolution in annealed d = 45 nm sample*

The observation of grain coarsening in extremely fine nanocrystalline metals during wear stands in stark contrast to reported observations of grain refinement in coarse-grained metals that leads to the development of nanocrystalline triobolayers [16, 34, 35, 55]. Argibay et al. [37] theorized that coarsening of fine-grained nanocrystalline microstructures and refinement of coarse-grained materials should converge to the same steady-state grain size, which they hypothesized defined the boundary between low and high friction. This work was more recently followed by Our results on the $d = 3$ nm sample above suggest that this steady-state grain size is perhaps defined by the cross-over from grain boundary-mediated to dislocation-based plasticity. To test if dislocation plasticity in a nanocrystalline alloy can create an evolved layer with this same grain size, an additional 1000 cycle scratch wear experiment was performed on the sample that had been annealed to give $d = 45$ nm. This sample therefore had the same composition but a different starting grain size, one above the proposed cross-over grain size. To study the microstructure, TEM lift-out was performed on the pre- and post-wear sample, with the results presented in Figure 7(a) and (b). Figure 7(b) shows that grain refinement is indeed occurring near the surface below the probe. The evolved layer near the surface is ~200 nm deep, which is similar to the thickness of the coarsened layer in the $d = 3$ nm sample. Figure 7(c) presents an SEM image of the wear track after 1000 cycles, where similar debris development and a sharp wear trench are observed.

Magnified views of the surface refinement are shown in Figure 8, where evidence of the development of many fine nanocrystals can be seen in both images. The average grain size in this layer was measured to be 9 nm, which is also similar to the 10 nm grain size found in the original sample. These findings help support the model proposed by Argibay et al. [38] since it seems that the surface layer converges to nearly the same average grain size for an identical level of applied surface stress. Moreover, the grain size distribution also spanned the range of 3 – 40 nm, indicating that both fine-grained ($d < 10$ nm) and coarser-grained (d > 10 nm) films of Ni-19 at.% W evolve to an identical structure. For the $d = 45$ nm sample, grain-boundary-mediated plasticity should be shut off and Figure 8 even shows evidence of mottled contrast inside many grains, providing evidence of dislocation plasticity. A major proposition in this area is that the steady-state grain size changes as a function of applied surface stress [38, 39]. Additional experimental evidence of this claim would help to validate and inform this model.

## 5. Conclusions

In this work, a systematic analysis of grain structure evolution during nanocrystalline wear was carried out on Ni-19 at.% W with an as-deposited grain size of 3 nm. Scratch wear experiments were performed using a conical diamond tip, with repeated back-and-forth wear tests from 1 – 1000 cycles being performed while tracking the volume of worn material at regular 10 cycle intervals. After analyzing the wear volume evolution, key cycling levels of interest were identified and characterization of the subsurface microstructure was performed to directly connecting wear rate to changes in microstructure. The following conclusions can be drawn from this study:

- Nanocrystalline metals are dynamic during wear, with microstructure evolving significantly below the sliding contact. Changes in wear rate can be directly correlated to changes in subsurface grain structure and the activation of different plasticity mechanisms.

- After the surface matches the counterbody and repeated plasticity occurs, a gradient nanostructure develops through grain boundary-mediated plasticity, with grain growth near the surface and a gradual transition to the bulk microstructure below. This gradient correlates with the expected stress field below the surface.

- Continued cycling allows for the extension of the grain growth deeper into the material, resulting in a clearly defined evolved layer and the disappearance of the gradient structure. The saturation in grain growth is attributed to the suppression of grain boundary plasticity mechanisms and a crossover to dislocation-based plasticity.

- Coarsened Ni-19 at.% W ($d$ = 45 nm) films exposed to identical wear testing show evidence of grain refinement. Interestingly, this refined layer reaches a nearly identical average grain size and grain size distribution, albeit through dislocation based plasticity. The observation of a common steady-state grain size provides evidence of the importance of the transition between different deformation mechanisms.

This study verifies that nanocrystalline grain structures act collectively to alter grain structure during the cyclic deformation imposed through nanocrystalline wear. This study identifies the intermediate gradient nanograined structure, which has not been observed previously, and connects different subsurface grain structures directly to instantaneous wear rates.

**Acknowledgements**

This research was supported by the National Science Foundation through the Division of Civil, Mechanical and Manufacturing Innovation under Award No. CMMI-1462717.

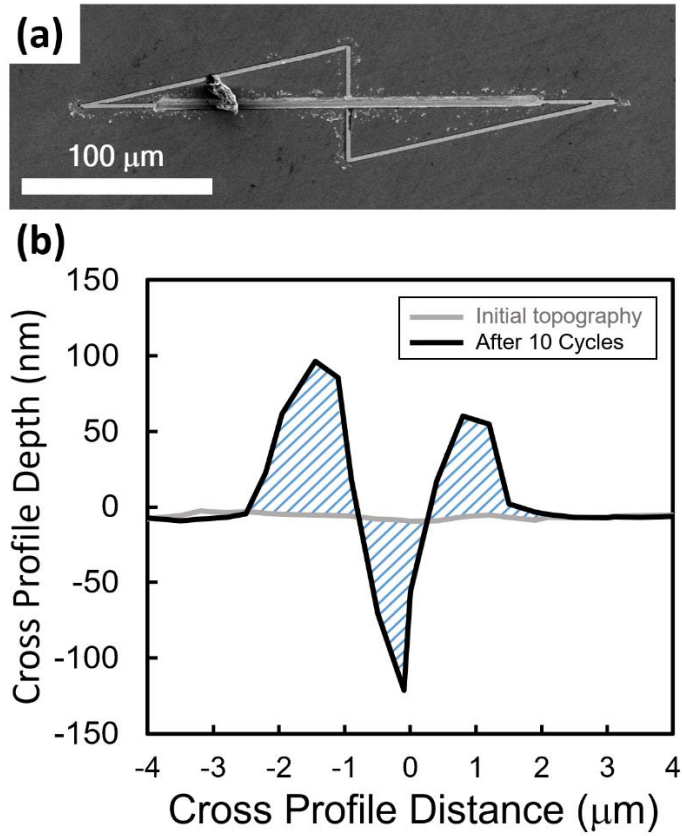

**Figure 1.** (a) An SEM image of a 1000 cycle scratch test that demonstrates the mechanics of the wear test experiment. (b) Calculation of the wear area is obtained by taking the integral of both the pileup and the scratch trench.

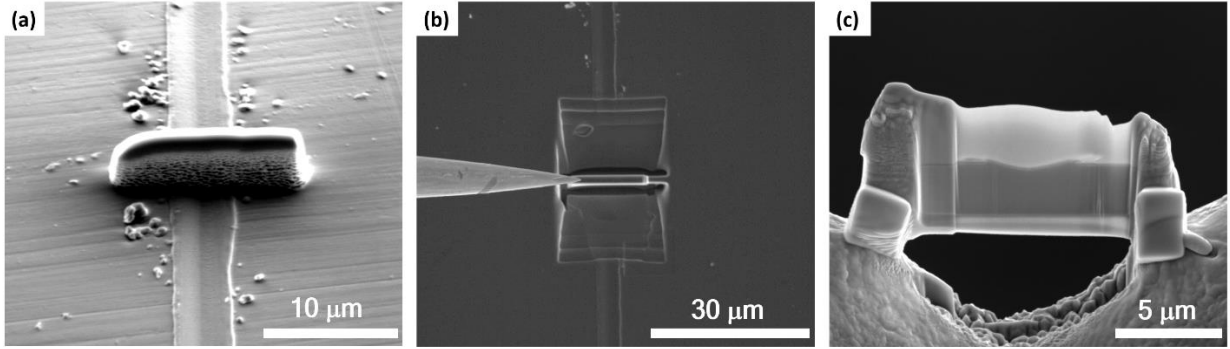

**Figure 2. (a) Deposition of a protective Pt layer over the wear track. (b) Removal of a cross-sectional lamella from the wear track. (c) TEM lamella before final thinning, where the entire cross-section of the scratch is visible under the Pt layer.**

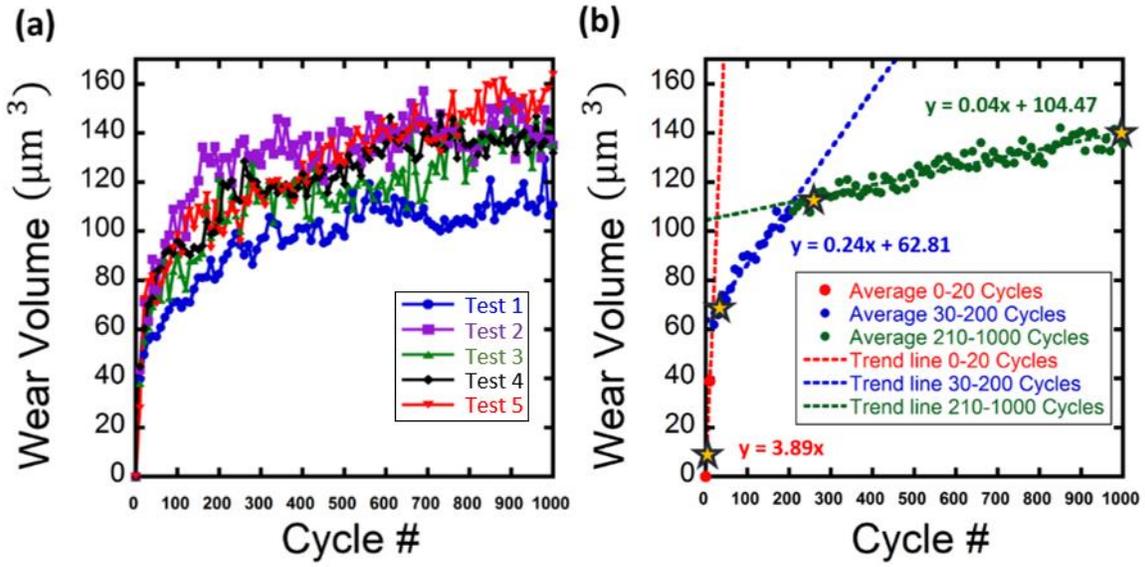

**Figure 3.** (a) Wear volume as a function of scratch cycle number for five separate 1000 cycle scratch experiments. (b) Average wear volume from the five tests, with linear regressions fit to the three distinct wear regimes. Gold stars indicate cycling levels where TEM characterization was performed.

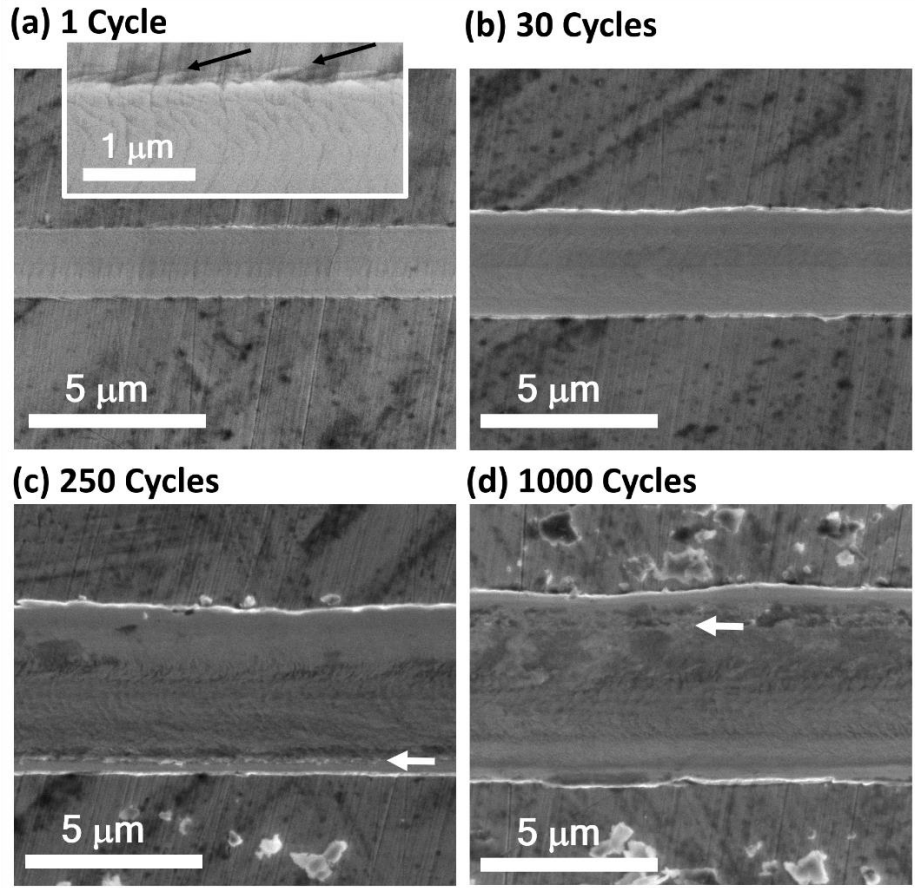

**Figure 4.** Scratch trench morphology after (a) 1 cycle, (b) 30 cycles, (c) 250 cycles, and (d) 1000 cycles. The inset to (a) shows evidence of strain localization near the edge of the wear track. Separation of the pileup is marked by white arrows in (c) and (d).

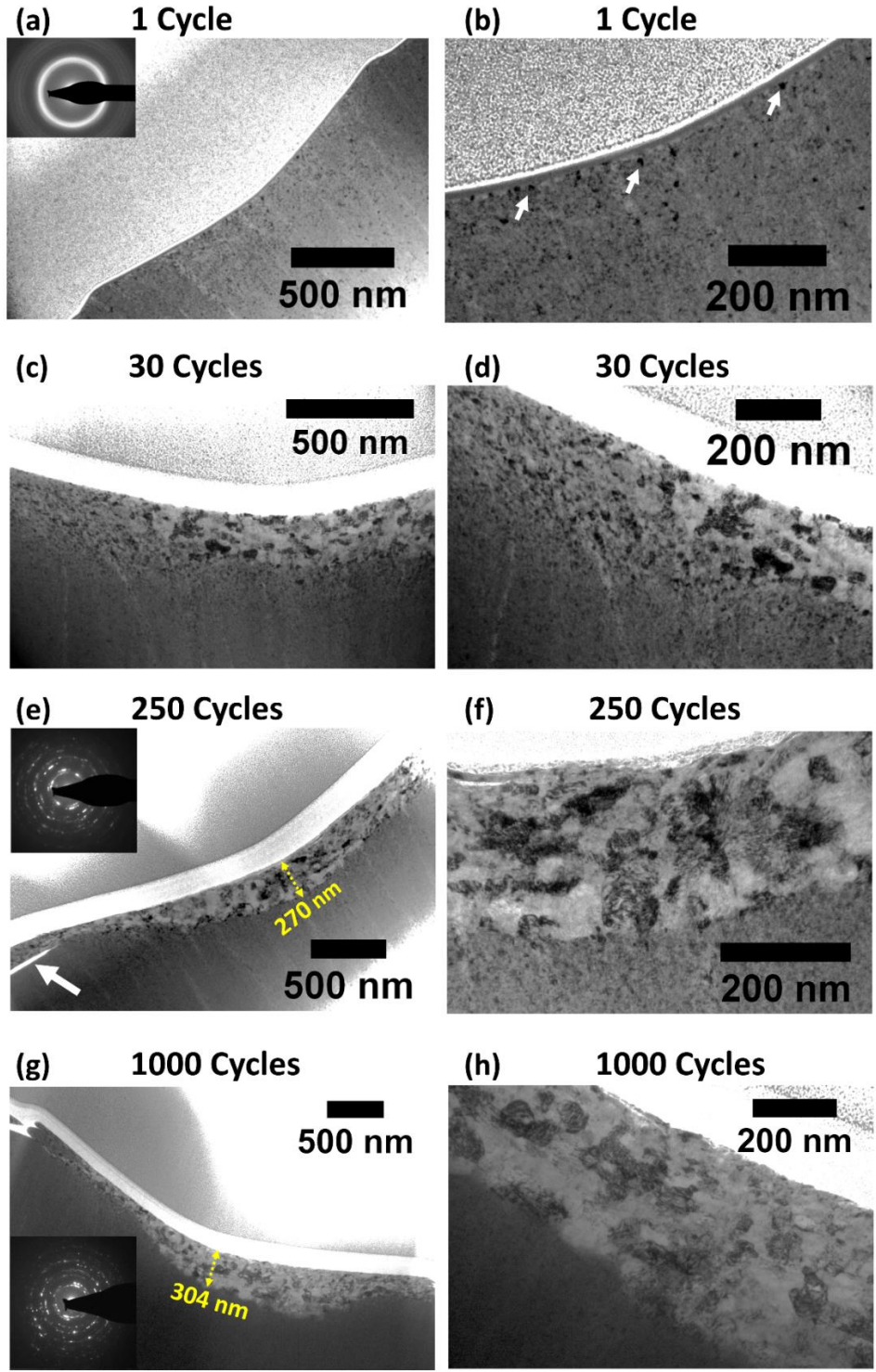

**Figure 5. TEM images showing near-surface microstructural evolution after 1 cycle, 30 cycles, 250 cycles, and 1000 cycles. Parts (a), (c), (e), and (g) show the overall evolution around the scratch tip, while parts (b), (d), (f), and (h) are magnified images of the microstructure near the center of the trench. An average grain size of 11 nm is found after 250 cycles, and an average grain size of 10 nm after 1000 cycles.**

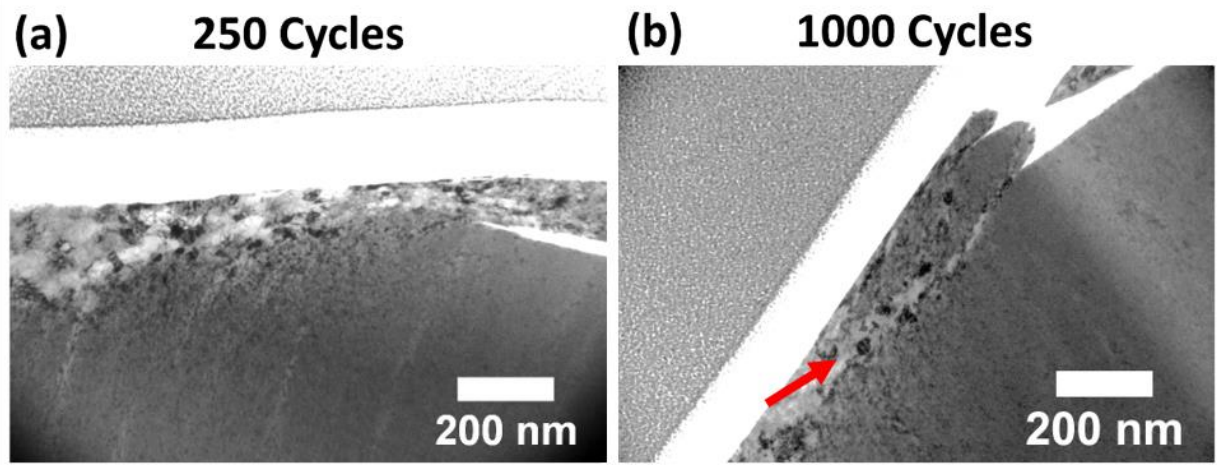

**Figure 6. TEM images at the trench edge for (a) 250 cycles and (b) 1000 cycles. A gradient structure can be found even at high cycling levels for the material near the edge which is deformed less. Pileup removal can be seen in both images as well as flow localization in the form of a shear band in image (b).**

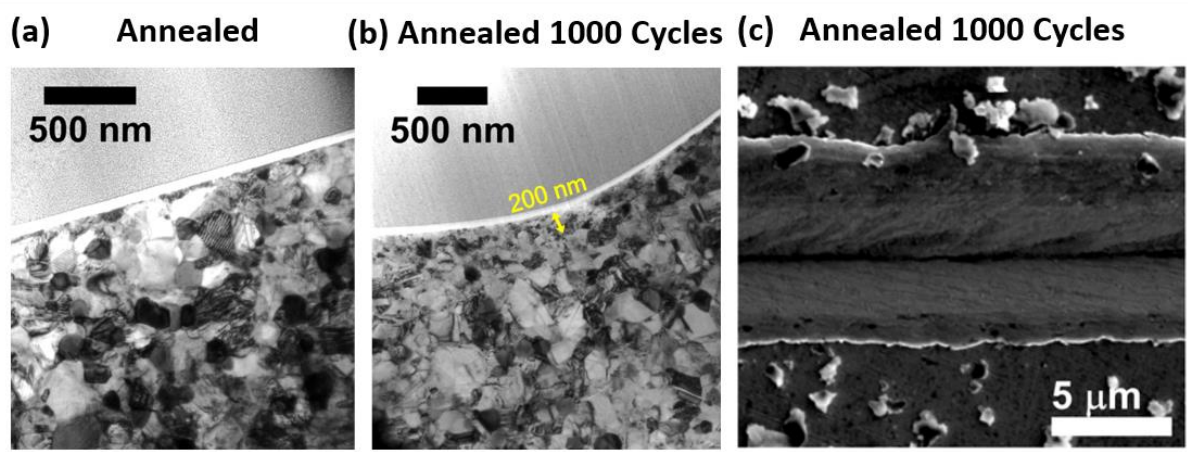

**Figure 7.** (a) TEM image of the unworn *d* = 45 nm film. (b) Post-worn cross-section after 1000 cycles indicating refinement below the surface. (c) SEM of the wear track after 1000 cycles.

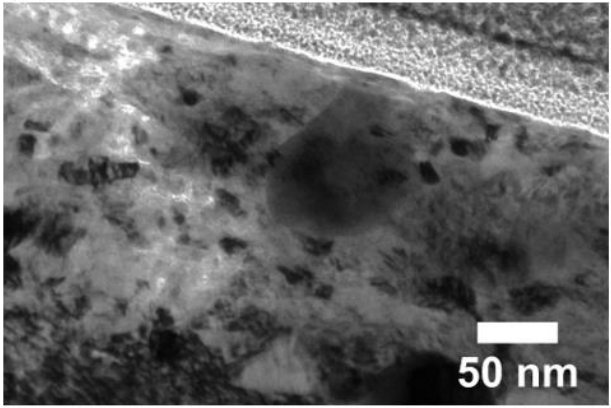 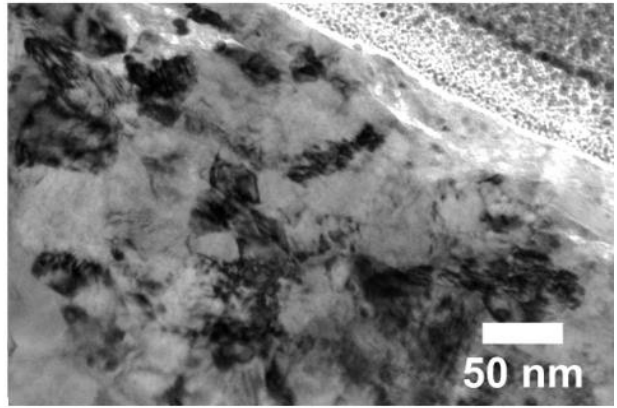

**Figure 8. Magnified TEM images of the evolved layer beneath the scratch tip. Clear refinement of the grain structure is observed. Detailed characterization of the layer gives an average grain size of 9 nm, similar to the grain size of the coarsened layer in the fine grained samples.**